\documentclass[fleqn,usenatbib]{mnras}
\usepackage{comment}
\usepackage{newtxtext,newtxmath}
\usepackage[T1]{fontenc}
\usepackage{ae,aecompl}
\usepackage{xcolor}

\usepackage{graphicx}	% Including figure files
\usepackage{amsmath}	% Advanced maths commands
\usepackage{amssymb}	% Extra maths symbols

\definecolor{darkred}{HTML}{E90052}
\title[The clustering of GRBs in the Hercules--\-Corona~Borealis Great Wall: the largest structure in the Universe?]{The clustering of gamma-ray bursts in the Hercules--\-Corona~Borealis Great Wall: the largest structure in the Universe?}

\author[I. Horv\'ath et al.]{
I. Horvath$^{1}$\thanks{E-mail: horvath.istvan@uni-nke.hu},
D. Sz\'ecsi$^{2}$, 
J. Hakkila$^{3}$, 
\'A. Szab\'o$^{4}$, 
I. I. Racz$^{1,5}$, 
L. V. T\'oth$^{6}$,\newauthor
S. Pinter$^{1}$ and
Z. Bagoly$^{5}$\\
\\
$^{1}$National University of Public Service, Budapest, Hungary\\
$^{2}$I. Physikalisches Institut, Universit\"at zu K\"oln, Z\"ulpicher-Str. 77, D-50937 Cologne, Germany\\
$^{3}$Department of Physics and Astronomy of College of Charleston, Charleston, SC, USA\\
$^{4}$Universit\"at Hamburg, Fachbereich Mathematik, Bundesstr. 55, D-20146 Hamburg\\
$^{5}$Department of Physics of Complex Systems the  E\"otv\"os Lor\'and University,	P\'azm\'any P\'eter s\'et\'any 1./A, 1117 Budapest, Hungary\\
$^{6}$Department of Astronomy of the  E\"otv\"os Lor\'and University,	P\'azm\'any P\'eter s\'et\'any 1./A, 1117 Budapest, Hungary\\
}

\date{Accepted XXX. Received 2020 May 16; in original form 2020 May 16}

\pubyear{2020}

\begin{document}
\label{firstpage}
\pagerange{\pageref{firstpage}--\pageref{lastpage}}
\maketitle
\providecommand*{\donothing}[1]{}
\begin{abstract}

The Hercules--Corona Borealis Great Wall is a statistically significant clustering of gamma-ray bursts around redshift 2. Motivated by recent theoretical results indicating that a maximal Universal structure size may indeed coincide with its estimated size (2$-$3\,Gpc), we reexamine the question of this Great Wall's existence from both observational and
theoretical perspectives. Our statistical analyses confirm the clustering's presence in the most reliable data set currently available, and we present a video showing what this data set looks like in~3D. Cosmological explanations (i.e. having to do with the distribution of gravitating matter) and astrophysical explanations (i.e. having to do with the rate of star formation over cosmic time and space) regarding the origin of such a structure are presented and briefly discussed and the role of observational bias is also discussed at length. This, together with the scientific importance of using gamma-ray bursts as unique cosmological probes, emphasises the need for future missions such as the THESEUS satellite which will provide us with unprecedentedly homogeneous data of gamma-ray bursts with measured redshifts. We conclude from all this that the Hercules--Corona Borealis Great Wall may indeed be the largest structure in the Universe -- but to be able to decide conclusively whether it actually exists, we need THESEUS.

\end{abstract}

\begin{keywords}
gamma-ray burst: general -- gamma-rays: general -- methods: data analysis -- methods: statistical -- 
Cosmology: large-scale structure of Universe -- 
Cosmology: observations 
%-- Cosmology: distance scale
\end{keywords}

%%%%%%%%%%%%%%%%%%%%%%%%%%%%%%%%%%%%%%%%%%%%%%%%%%

%%%%%%%%%%%%%%%%% BODY OF PAPER %%%%%%%%%%%%%%%%%%

\section{Introduction}

The largest observed structure in the Universe so far reported was inferred from a clustering of gamma-ray bursts (GRBs). This clustering, called the {\it Hercules--Corona Borealis Great Wall} \citep{hhb14,hbht15}, is characterised by an anisotropy in the GRB angular distribution at a redshift ($z$) of around~2, and is found by statistical analyses of the GRB angular locations and assumed radial distances. The clustering's celestial position corresponds to the constellations of Hercules and Corona Borealis, hence the name. 

Although the clustering's initial discovery \citep{hhb14} was supported by subsequent observations \citep{hbht15}, the physical nature of the structure is still unknown. 
Because of this, it is important to continue verifying the cluster's existence as the GRB database grows. In this paper we examine this question further by analysing an even larger, yet reliable data set. 
Our motivation to do so comes from the recent theoretical results of \citet{Canay:2020}, who find that the screening length $\lambda$ above which large-scale structure formation is suppressed do actually coincide with the size of the Hercules--Corona Borealis Great Wall (2.6~Gpc). However, for a complete picture alternative explanations should also be discussed, including the possibility of this structure being an observational artefact \citep{Ukwatta:2016}.
 
The angular distribution of GRBs has been
extensively studied over the past few decades. 
For the most part, 
GRBs have been found to be uniformly distributed on the sky 
\citep{Briggs96,bal98,bal99,mesz00,mgc03,vbh08,2019MNRAS.486.3027Ripa,2019MNRAS.490.4481And},
although some subsamples appear to deviate from isotropy
\citep{bal98,Cline99,mesz00,li01,mgc03,vbh08}. Apart from the Hercules--Corona Borealis Great Wall identified by our group \citep{hhb14,hbht15}, another large GRB-defined structure has been recently reported. This structure, found by \citet{Balazs:2015} in the redshift range of $0.78 < z < 0.86$ using a variety of statistical methods, is called the \textit{Giant GRB Ring}. It appears to be somewhat smaller (1.72~Gpc) than the Hercules--Corona Borealis Great Wall (2$-$3~Gpc), but its presence in the data has been confirmed by further, elaborate statistical analysis \citep{Balazs:2018}.
While the physical and astrophysical nature of both structures are not conclusively understood \citep[cf. the discussions in][]{Balazs:2015,Balazs:2018}, their existence may provide a challenge to standard assumptions about universal homogeneity and isotropy (i.e. the cosmological principle; see however \citealt{Li2015}). This means that studying structures like these is of high scientific importance.

Aside from GRBs, there are other large structures discovered in the 21st century. Namely, 
the Sloan Great Wall which is a giant filament of galaxies has a size of $\sim$\,0.4~Gpc \citep{Gott05}, while the Huge Large Quasar Group which possibly consist of 73~quasars has a size of 1.2~Gpc \citep{clo12}. Nonetheless, both GRB-defined structures (the Giant GRB Ring and the Hercules--Corona Borealis Great Wall) exceed these in size.

Here we focus our attention to the Hercules--Corona Borealis Great Wall (from now on, Great Wall). We aim to reexamine the question of its existence from various aspects, and offer both arguments and counter-arguments for it. In addition to performing statistical analyses on an up-to-date and reliable data set and discussing recent results which challenged the existence of the Great Wall \citep{2020MNRASCS}, we also offer various scenarios for the origin of such a large structure. This makes the present work an organic completion to the first two papers on the Great Wall \citep{hhb14,hbht15} in which formation scenarios were only passingly mentioned.
Additionally, we present a video showing an orthographic 3D representation of the `GRB Universe', with the Great Wall marked. And last but not least, we offer counter-arguments too: by discussing the relevance of observational bias in the collecting of the data and guided by the relevant investigation of \citet{Ukwatta:2016}, we conclude that new, more homogeneous data sets may be required to draw conclusive answers about the anisotropy of the Universe and the existence of large structures. We suggest that the proposed satellite mission THESEUS is the most ideal instrument to collect such data. 

The paper is organised as follows. Section~\ref{sec:data} describes the data set we use here, and presents the video of the GRB Universe. Section~\ref{sec:analyses} discusses the statistical analyses we perform on the data (namely, a nearest-neighbour analysis and a point-radius bootstrap method). Section~\ref{sec:comments} addresses techniques applied and results obtained in previous work. Section~\ref{sec:discussion} offers theoretically oriented discussions about the possible cosmological and/or astrophysical origin of such a large structure as the Great Wall. Amongst these, Section~\ref{sec:caveats} in particular discusses the role of observational bias and argues for the uttermost importance of such a future mission as the THESEUS satellite. Section~\ref{sec:summary} summarises our arguments both for and against the existence of the Hercules–Corona Borealis Great Wall being amongst the largest structures in the Universe.

\section{The data}\label{sec:data}

\subsection{Previous work and the current sample}

The first identification of a clustering of GRBs in \cite{hhb14} 
was based on an analysis of 283~GRBs that had been 
localised prior to July 2012.
By November 2013, the redshifts of 361 GRBs had been 
determined, which represented a 28\% increase in the sample size.
In \citet{hbht15} therefore this sample has been studied:  indeed,
the number of GRBs around redshift $z=2$ ($1.6 \le z < 2.1$)
where the clustering resides had increased from 31 bursts
to 44 bursts in that 1.5~years' time, a 42\% sample size increase that was large enough
to warrant the updated analysis.

Based on this data set, \citet{Balazs:2015} also found the Great Wall using a statistical method that was independently developed from those in \citet{hhb14,hbht15} and the present work. Although these authors focused on reporting another large structure (the Giant GRB Ring), their Fig.~2 shows the Great Wall too, with a 1$\sigma$\,$-$\,2$\sigma$ significance.

As of March 2018,
the redshifts of 487 GRBs have been measured (mostly comprised of bursts detected by NASA's
Swift experiment)\footnote{\url{http://www.astro.caltech.edu/grbox/grbox.php} 
}.
This represents an increase in sample size of yet another 35\% in the five-year period since the \cite{hbht15} analysis. 
This largest sample is what we work with and analyse here. 

\subsection{On the importance of data selection}\label{sec:Christiandata}

Recently \citet{2020MNRASCS} used a data set\footnote{\url{http://www.mpe.mpg.de/~jcg/grbgen.html} } containing almost every GRB observed by any instrument, hereafter referred to as the ``Greiner table''. One of the goals of this data set is the support of the fast precise localization and optical afterglow measurements of the GRBs, therefore this table is more of a subjective data collection than a formal database. It includes entries with all published redshift estimations (e.g. optical afterglow, photo-z, Lyman-$\alpha$ limits) which could support the observational work. As explained below, use of the table without carefully accounting for all included redshift entries is considered ill-advised.

We have downloaded the GitHub files\footnote{ \url{https://github.com/Sam-
2727/Gamma-ray-burst-isotropy}} used by \citet{2020MNRASCS} and have compared them to the Greiner table from which they were taken. We notice, first of all, that the GitHub files 
failed to exclude 25 GRBs having photometric redshifts with large uncertainties as well as 24 GRBs marked by question marks. Second, more than a dozen GRBs in the Greiner table are marked as having redshifts differing by more than 0.2 (some by as much as 0.4) from the values given in the notes of the Gamma-ray Coordination Network\footnote{\url{http://gcn.gsfc.nasa.gov/}}. Thus Christian's database contains a significant amount of data characterized by potentially large uncertainties. Only when correcting for these values or at least properly discussing their effects can the results and conclusions of \citet{2020MNRASCS} be considered fully reliable.

Our catalog database (published by the Caltech Astronomy Department) tries to avoid the poorly-defined measurements found in the Greiner table.  This database has been carefully reviewed by its creators to exclude measurements with large systematic uncertainties, which has resulted in a small yet homogeneous catalog filled with data that can be more easily trusted for spatial analysis.

\begin{figure}
\href{https://hera.ph1.uni-koeln.de/~szecsi/Publications/Videos/anim-Hercules.mp4}{\includegraphics[width=\columnwidth]{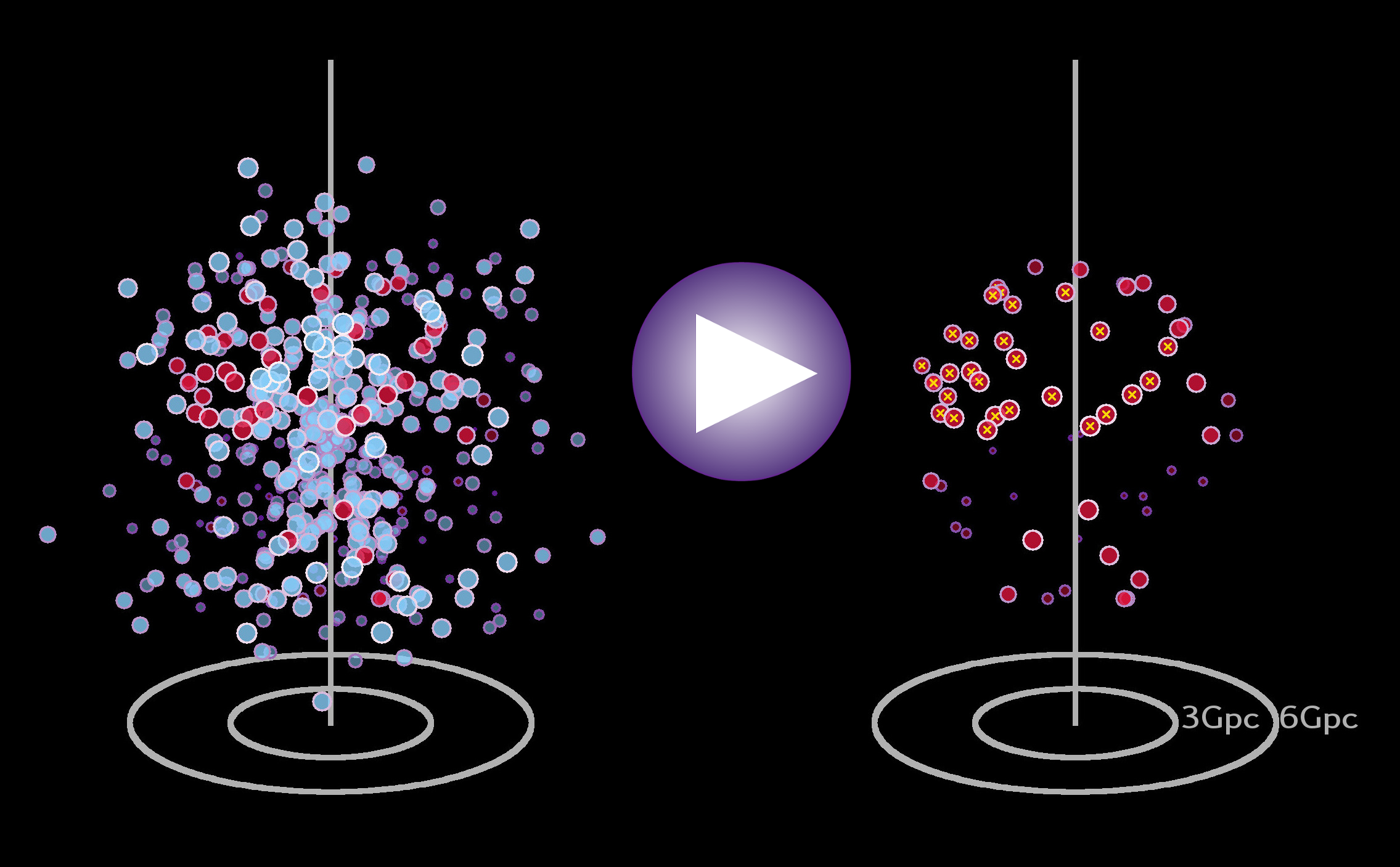}}
\caption{
    \textit{The GRB Universe.} Orthographic 3D~representation of our 4D~Universe as seen by GRBs in our data set (details in Sect.~\ref{sec:GRBUniverse}). \textit{Left:} all GRBs with known distances (blue). Those between 1.6~$<z<$~2.1 are marked (red). The size of the dots, as well as their shading, refer to the distance of the given object from an imaginary external viewer. 
    \textit{Right:} only GRBs in the redshift range 1.6~$<z<$~2.1 are shown, with those that likely belong to the Great Wall marked by yellow crosses (i.e. with galactic coordinates $b \gtrsim 0$ and $0 \lesssim l \lesssim 180$). 
    Comoving distances, converted from redshift $z$ values, are indicated at 3~Gpc and 6~Gpc. A video showing the same \textit{GRB Universe} rotating around, can be seen at or downloaded \href{http://hera.ph1.uni-koeln.de/~szecsi/Publications/Videos/anim-Hercules.mp4}{under this link}. 
}\label{fig:3D}
\end{figure}

\subsection{The GRB Universe -- a video}\label{sec:GRBUniverse}

We believe in the power of visualisation. Therefore, together with presenting statistical analyses on the data in Sect.~\ref{sec:analyses}, we present a video showing a 3D representation of the data set (rotating around in various directions). Figure~\ref{fig:3D} shows only a snapshot, and we encourage the reader to take a look at the whole video \href{https://hera.ph1.uni-koeln.de/~szecsi/Publications/Videos/anim-Hercules.mp4}{under this link}. 

The video represents our 4~dimensional Universe only in 3~dimensions. Every GRB in the sample is shown by one dot, according to its redshift and observed celestial position. To calculate orthographic coordinates in parsecs, redshift values were converted into comoving distances using standard cosmological parameters \citep{Planck:2018}. This means that every spherical (i.e. 2~dimensional) layer is actually a 3~dimensional space at a given cosmic age (where time is given by $z$ or, as in here, the comoving distance).

In the video, the Earth-based observer is located in the centre of the distribution. The viewer of the video is, however, sitting `outside' of the Universe and looking at it from an imaginary 10\,Gpc distance. The size and brightness of the dots are both scaled with their respective distance from the viewer, as one over distance squared; that is, the smaller and dimmer a dot, the farther away from the viewer it is. 

In the video, the Great Wall's approximate position is marked such that its redshift range is given in red and its galactic coordinates are given in yellow. Note however that we can gladly provide videos of the same style for GRBs in any redshift range and galactic locations upon request.

\section{Statistical analyses}\label{sec:analyses}

\subsection{Nearest-neighbour analysis}\label{sec:nearest}

The GRB cluster representing the Great Wall was previously localised in the $1.6 < z < 2.1$ redshift range. We continue studying GRBs in this approximate redshift range while maintaining independent radial sub-samples of similar sizes. 
Amongst the 487 GRBs in our current sample there are 64 in the $1.6 < z < 2.1$ redshift range (13.14\%). Thus, the sub-sample size chosen for this analysis is 64.
(Our initial analysis, \citealp{hhb14}, had 31 of 283 GRBs (10.95\%) in this redshift range, and our second analysis, \citealp{hbht15}, had 44 of 361 GRBs (12.19\%).)

The first statistical tool we apply to the sample is the nearest-neighbour analysis and the second (presented in Sect.~\ref{sec:point}) is the point-radius bootstrap method. As mentioned previously, we subdivide our sample into equal-sized radial groups, with each group containing 64 GRBs. Thus, we have six non-overlapping groups within $z$ ranges defined in Table~\ref{tab:tablez}. The nearest 51 ($z < 0.45$) and the farthest 52 ($3.65 < z$) GRBs have been omitted.  

\begin{table}
	\centering
	\caption{The equal-sized (64 GRBs in each) radial groups.}
	\label{tab:tablez}
	\begin{tabular}{ c | c | c | c | c | c | c | c | c }
	    group name & GR1 & GR2 & GR3 & GR4 & GR5 & GR6  \\ \hline
        $z_{min}$  & 0.45 & 0.82 & 1.21 & 1.6 & 2.1 & 2.67  \\ \hline
       $z_{max}$ & 0.82 & 1.21 & 1.6 & 2.1 & 2.67 & 3.65  \\ 

	\end{tabular}
\end{table}

The angular distribution of each radial group can be studied by independently applying the {\it k}th nearest-neighbour
statistics to the data points (i.e. GRB celestial positions) in each group. 
Nearest-neighbour statistics have been often used in the literature  \citep{1989MNRAS109S,1997ApSSMS,2017MNRASTarnop} to characterise isotropy. The nearest-neighbour distance indicates the angular separation between each GRB and its closest neighbour. Similarly, the second-nearest neighbour indicates the angular separation between each GRB and its second-closest neighbour. The total of all {\it k}th nearest-neighbour defines a set that describes a distribution's angular characteristics almost completely.

We perform Monte Carlo simulations for our GRB sample to characterise nearest-neighbour distributions and their deviations from isotropy. We generate $10\,000$ catalogues using 64 GRB locations selected randomly from the observations, without repetition.
We measure all the {\it k}th nearest-neighbour distances (where $k \le 63$) from these catalogues, and use them to create the empirical reference function. As every catalogue has its own distance points, we create a joint equidistant $0-\pi$ angular scale onto which each catalogue is extrapolated. The mean of these extrapolated distance functions from the catalogues is used as an empirical reference function. 
This way the reference contains all the effects coming from the uneven sky exposure as the catalogues contain all the angular distribution information. The suspected anisotropy is also included in the catalogs, which lowers the sensitivity of the test.  However we suspect that this effect is small, considering the weakness of the suspected anisotropy signal.  
For the statistics, similarly to the Kolmogorov--Smirnov distance $D$ we calculate the largest  difference between each catalogue's {\it k}th nearest-neighbour distributions and this reference function.

The same approach is used for the real GRB data too: the $D$-bvalue distributions are determined for every {\it k}th nearest-neighbour for the real GRB groups \citep{racz2017AIP}. Thus, the number of random catalogues containing $D$ distances larger than that of the real group provides an estimate of how anisotropic the real group is. This can be used to approximate the probability in hypothesis testing.

Applying this technique on our six radial GRB groups, we find that Group~4 (GR4) -- the one previously found to contain the Great Wall -- is 
the most anisotropic. (The second most anisotropic is the neighbouring GR5.) For almost every {\it k}, GR4 ($1.6 < z < 2.1$ ) contains the largest values of $D$. 
There are several cases for GR4 in the $k=6-20$ range where the real data's $D$ distance is greater than that found for 99.75\% ($2.5 \sigma$ of the normal distribution) of the random ones, however, no particularly strong peak is found. There is no significant grouping of the larger $D$ values compared with the random distributions.
This result hints that a more sensitive statistical measure should be used to check the viability of this anomaly because maximum distance methods like the Kolmogorov--Smirnov method are not very sensitive (although they are robust). For example, if the sky exposure could be determined (or assumed, cf. Sect.~\ref{sec:skyexposure}) the real spatial two-point correlation function could be applied \citep{SWIFT10..060,IAU15,Li2015} as a validation technique.

\begin{comment}

\begin{figure}
	\includegraphics[width=\columnwidth]{nn64sz24.png}
   \caption{24th nearest-neighbour distributions for  redshift-limited GRB samples. WE DO NOT NEED THIS IT WILL BE DELETED SOON
 as u see the yellow is the $1.6 < z \le 2.1$ however the dark blue $2.1 < z < 2.7$ group has the biggest deviation as I write in the text)}
  \label{fig:fig2}
\end{figure}

\end{comment}

\subsection{The point radius bootstrap method}\label{sec:point}

In order to test for the existence of the clustering in the current database while accounting for known sampling biases, we apply the point radius bootstrap method.
This method was
described in Sec.~5 of \citet{hhb14}. We repeat the same analysis here, albeit for a much larger data set (as explained in Sect.~\ref{sec:data}).

To use the point radius bootstrap method, we
assume that the sky exposure is independent of~$z$ (also cf. Sect.~\ref{sec:caveats} on the caveats).
We randomly choose a sub-sample of 64~GRBs from the observed data-set. (We have tested that the results do not depend on including or excluding GRBs between $1.6 < z \le 2.1$ in this sub-sample.)
Then we compare the sky distribution of
this sub-sample with the sky distribution
of 64~GRBs with $1.6 < z \le 2.1$. 

To study the selected GRBs in two dimensions,
we choose random locations on the
celestial sphere and determine how many of the
64~points lie within a circle of predefined angular radius (for example, within $20^{\circ}$).
We repeat the process a large number of times, i.e. 20\,000 times. 
From these 20\,000 cases we select the largest number of GRBs found within the angular circle (for more details about this method see our previous works, \citealt{hhb14,hbht15}, where the same method was applied).

	% Table 1
\begin{table}
	\centering
	\caption{Largest number of GRBs found within a certain area of the sky.}
	\label{tab:table1}
	\begin{tabular}{ c | c | c | c | c | c | c | c | c }
	    radius & $32.9^\circ$ & $34.9^\circ$ & $36.9^\circ$ & $38.7^\circ$ & $40.5^\circ$ & $42.3^\circ$ & $43.9^\circ$ \\ \hline
        surf. area  & 0.08 & 0.09 & 0.10 & 0.11 & 0.12 & 0.13 & 0.14 \\ \hline
        GRBs & 21 & 23 & 24 & 27 & 28 & 30 & 32 \\ 
 &  &  &  &  &  &    \\ \hline
 	    radius & $45.6^\circ$ & $47.2^\circ$ & $48.7^\circ$ & $50.2^\circ$ & $51.7^\circ$ & $53.1^\circ$ & $54.5^\circ$  \\ \hline
        surf. area  & 0.15  & 0.16 & 0.17 & 0.18 & 0.19 & 0.20 & 0.21 \\ \hline
         GRBs & 33 & 33 & 33 & 34 & 35 & 36 & 36 \\ 
 &  &  &  &  &  &    \\ \hline
	    radius & $55.9^\circ$ & $57.3^\circ$ & $58.7^\circ$ & $60.0^\circ$ & $61.3^\circ$ & $62.6^\circ$ & $63.9^\circ$ \\ \hline
        surf. area  & 0.22 & 0.23 & 0.24 & 0.25 & 0.26 & 0.27 & 0.28 \\ \hline
         GRBs & 37 & 37 & 37 & 38 & 38 & 38 & 39 \\ 
	\end{tabular}
\end{table}

%--------------------------

\begin{figure}
	\includegraphics[width=\columnwidth]{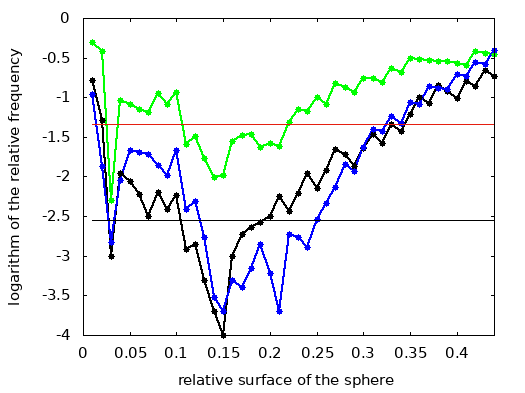}
    \caption{Results of the Monte Carlo bootstrap point-radius test on a variety of different angular scales. 
The horizontal coordinate is the area of the circle in the sky 
relative to the whole sky ($4\pi $). The vertical
coordinate is the logarithm of the relative frequency 
found from the 10,000~runs. 
The calculations were made for 64 GRBs in the $1.6 < z < 2.1$ range (black), for the 77 GRBs  in the $1.6 < z < 2.3$ range (blue) and the 77 GRBs in the $1.5 < z < 2.1$ range (green).
Horizontal red (black) line shows the $2\sigma$ 
($3\sigma$) deviations.
}
    \label{fig:figMCBP}
\end{figure}

This analysis is performed with the
64~GRBs that belong to our location of interest, and also with 64~randomly chosen 
GRB locations from the observed data (i.e. from the known 487 GRBs).
We repeat the experiment 10\,000~times 
in order to understand the statistical variations of this sub-sample.
We also perform the same technique using angular circles
of different radii. 
The frequencies obtained this way are shown in Figure~\ref{fig:figMCBP} (black).

Table~\ref{tab:table1} shows the maximum number of GRBs in a given angular circle.
The most significant deviation from isotropy appears in a circle covering 15 percent of the sky (see Fig.~\ref{fig:figMCBP}); at least 33 GRBs are contained inside this circle.
The significance reaches $3\sigma$ between regions covering 11 percent and 20 percent of the sky (Fig.~\ref{fig:figMCBP}, the black horizontal line shows the $3\sigma$ limit). In these regions, between 27 and 36 GRBs are found (out of 64).

We check whether the sky distribution anisotropy spans a larger $z$ interval than the $1.6 < z \le 2.1$ range in which the clustering was originally identified. To do this, we apply the point radius bootstrap method to regions extended to other redshifts. We consider two such regions: one extended to smaller redshifts ($1.5 < z \le 2.1$) and the other extended to larger redshifts ($1.6 < z \le 2.3$). Since both volumes contain 77 GRBs, we choose a sub-sample of 77~GRBs from the observed dataset. Then we select random locations on the
celestial sphere and determine how many of the
77~points lie within a circle of predefined angular radius. 
We estimate statistics for this test by 
repeating the process 20\,000~times. 
From these 20\,000 Monte Carlo runs we select the largest number of GRBs found within the angular circle.
We repeat the process with 77~different randomly
chosen GRB positions (from the known 487 GRBs), and we repeat the experiment 10\,000~times 
in order to understand the statistical variations of this sub-sample.
We also perform the same technique using angular circles
of different radii. 

The result with the 77 GRBs from the $1.5 < z \le 2.1$ interval and 
with the 77 GRBs from the $1.6 < z \le 2.3$ interval are presented in Fig.~\ref{fig:figMCBP}.

Figure~\ref{fig:figMCBP} shows that the extended $z$ interval $1.5 < z \le 2.1$ (with 77 GRBs) contains a much less significant anisotropy than the one found in $1.6 < z \le 2.1$, and it never reaches the $3\sigma$ level (green). However, the extended $z$ interval $1.6 < z \le 2.3$ (again with 77 GRBs) shows a similar anisotropy at a similar significance level (blue in Fig.~\ref{fig:figMCBP}). For relative surface areas between 0.05 and 0.1, the 64 GRBs in the originally-defined $z$ interval $1.6 < z \le 2.1$ show the largest significance, but in the 0.17 - 0.27 interval, the 77 GRBs ($1.6 < z \le 2.3$) exhibit the greater significance. In both cases the minimum frequency is around 0.15 (containing 33 of 64 and 37 of 77 GRBs, respectively). The probability of finding 33 of 64 randomly-distributed GRBs in only $15\%$ of the sky is extremely low ($\approx 10^{-12}$). Similarly, the probability of finding 37 of 77 randomly-distributed GRBs in such a small fraction of the sky is also extremely low ($\approx 10^{-12}$). 

These results imply that \textit{the clustering of GRBs in the Hercules--Corona Borealis Great Wall is indeed statistically significant in the most} reliable \textit{database currently available}.
Note however that the assumption of randomness may not be entirely valid due to the anisotropic presence of galactic dust. We discuss possible caveats and observational biases in Sect.~\ref{sec:caveats}.

\section{Comments on techniques used in previous work}\label{sec:comments}

\subsection{On the inadequacy of applying the 2D Kolmogorov--Smirnov test to data on a sphere}\label{sec:2DKS}

Recently, \citet{2020MNRASCS} investigated the Great Wall's existence by applying the two-dimensional Kolmogorov--Smirnov test to a GRB data set that we discussed and found wanting in Sect.~\ref{sec:Christiandata}. They concluded that the test indicated only a 0.054 probability in support of the GRB clustering's existence. Furthermore, they raised some concerns about why \citet{hhb14} had used this test in the original paper discovering the GRB clustering but not in the follow-up analysis in \citet{hbht15}. Their question suggested that the 2D Kolmogorov--Smirnov test had been excluded from the latter analysis because it did not support the high confidence claimed by \citet{hbht15} about the existence of the clustering. We address these concerns here.

The Kolmogorov--Smirnov test is typically used for real-valued random variables since the unique structure of the real line~$\mathbb{R}$ allows for an unambiguous definition of the Kolmogorov--Smirnov distance. This distance of cumulative distribution functions is invariant under postcomposition by homeomorphisms of~$\mathbb{R}$. However, for random variables with values in other topological spaces, the definition cannot be extended in an invariant way in general. Nonetheless, several analogies of the Kolmogorov--Smirnov test have been proposed, cf. \citet{Peacock83} and later \citet{1987MNRASFF} who work with rectangular regions of the plane~$\mathbb{R}^2$. In our earlier work (\citealp{hhb14}), we implicitly assumed that this test could be adequately applied to two-dimensional sky distributions, i.e. to random variables with values in the sphere~$S^2$. However, we later realized that the method used by \citet{Peacock83} and \citet{1987MNRASFF} cannot be transported to the whole of~$S^2$ due to topological reasons. Of course, one can work with a punctured sphere by using a coordinate chart but the result of such an analysis will depend on the chosen coordinates. A further drawback of this (pseudo) Kolmogorov--Smirnov method is that it is not invariant under~$SO(3)$ transformations, i.e. simple rotations. A valid, $SO(3)$-invariant 2D Kolmogorov--Smirnov analysis on~$S^2$ is an interesting problem but it is out of scope here. 

For completeness, and to check our results against those of \citet{2020MNRASCS}, we have applied the (pseudo) 2D Kolmogorov--Smirnov test here to our present data set. Since there are 64 GRBs in the $1.6 < z < 2.1$ range (GR4), for comparison we choose redshift groups with 64 GRBs in each. There are 243 GRBs with $ z < 1.6$ and 180 GRBs with $2.1 < z $, allowing only three 64-GRB disjoint groups (GR1, GR2, GR3) at lower redshift and two (GR5, GR6) at higher redshift than that of GR4 (Table~\ref{tab:tablez} shows their redshift range). Table~\ref{tab:tableKST} shows the largest 2D Kolmogorov--Smirnov distances found between the groups, using the 2D Kolmogorov--Smirnov test as described in \citet{hhb14}. Note that the three largest Kolmogorov--Smirnov distances, corresponding to the greatest differences in the sky distribution, do relate to GR4 ($1.6 < z < 2.1$). We estimate the significance of these anisotropies by selecting 64 GRBs at random from the complete 487-GRB sample. The 2D Kolmogorov--Smirnov values are obtained by comparing these distributions, and the process is repeated 3999 times to generate a statistical estimate. Only 6.9\% of the runs produce Kolmogorov--Smirnov distances greater than or equal to 19 and only 4.1\% of the runs produce Kolmogorov--Smirnov distances greater than or equal to 20. These results are similar to the 5.4\% probability obtained by \citet{2020MNRASCS}. However, as mentioned previously, we discount these results because we believe that there are inherent flaws when applying the flat 2D Kolmogorov--Smirnov process on a sphere. 

\begin{table}
	\centering
	\caption{The 2D KST distances between the GRB groups.}
	\label{tab:tableKST}
	\begin{tabular}{ c | c |  c | c | c | c | c | c }
	    group name &  GR2 & GR3 & GR4 & GR5 & GR6  \\ \hline
        GR1  &   18 & 12 & {\bf19} & 17 & 14  \\ \hline
      GR2 &  &   16 & 14 & 18 & 16  \\ \hline
      GR3 &   &  & {\bf19} & 13 & 15  \\ \hline
            GR4   &  &  &  & 16 & {\bf20}  \\ \hline
                  GR5 &    &  &  &  & 13  \\ \hline
	\end{tabular}
\end{table}

\subsection{Addressing additional criticisms by \citet{2020MNRASCS}}\label{sec:C2}

\citet{2020MNRASCS} questioned the results of \citet{hhb14} obtained with the point-radius bootstrap method, criticizing the authors for assuming that the number of bursts within a random angular radius followed a binomial distribution. This unfounded criticism was due to an unfortunate misreading of \citet{hhb14}: binomial probabilities  were only included in a bracketed note to demonstrate extreme behavior, and this note was never used as the basis for a statistical argument. In fact, \citet{hhb14} stated that the distribution was selected to coincide with the sky exposure ({\it e.g.,} see Sect.~\ref{sec:skyexposure} in this paper), thus indicating that burst locations on the plane of the sky were not distributed randomly and could not be characterized by binomial probabilities. In attempting to correct this suspected error, \citet{2020MNRASCS} repeated the analysis with the point-radius bootstrap method using the \citet{hhb14} data and found p $\approx$ 0.002. This result is consistent with the \citet{hhb14} results (see their Fig.\,2) which found that the probability reached the $3\sigma$ level (p=0.0028) twice. Thus \citet{2020MNRASCS}'s result from the point-radius bootstrap method supports the conclusions of \citet{hhb14} instead of contradicting them.

In Sect.~3.3 of their paper, \citet{2020MNRASCS} concluded that \citet{hhb14} had overstated the significance of a GRB anisotropy between $1.6 < z < 2.1$. Applying the nearest-neighbour analysis to randomly-generated distributions, they found that ``181 out of the 5,000 samples had a higher K-statistic'' than the anisotropic distribution did. We are unfortunately unable to reproduce \citet{2020MNRASCS}'s nearest-neighbour result using data and programs downloaded from their GitHub files (cf. Sect.~\ref{sec:Christiandata}). When we run 5,000 random samples we only obtain a higher K-statistic 72 times rather than 181 times as quoted in the paper. Concerned that we might have obtained this result by statistical fluke, we repeated this 5,000-sample analysis 400 times. On average, we only obtained higher K-statistics 70.09 times out of 5,000 (with a standard deviation of 12.85, a minimum of 38 times, and a maximum of 98 times). Therefore, rather than the p-value of 0.0362 obtained by \citet{2020MNRASCS}, we find that p is 0.014 using the same data and program files. Again, these results support our hypothesis that the GRB anisotropy is significant as opposed to the conclusion stated by \citet{2020MNRASCS}.

\subsection{Are sampling biases responsible for the anisotropy?}\label{sec:ukwatta}

\citet{Ukwatta:2016} claimed that the GRB anisotropy found between $1.6 < z < 2.1$ \citep{hhb14} could be produced solely by anisotropies in the sample's sky exposure. To test this hypothesis, they generated three density maps. The maps correspond to a uniform sky exposure, Swift's sky exposure (taken from Swift's recorded instrument pointings; their Fig.\,4), and a combined exposure that was meant to convolve Swift's sky exposure with effects of interstellar extinction. However, the combined exposure, meant to most clearly represent the sampling biases, appears to be invalid for several reasons. 

First, \citet{Ukwatta:2016} constructed the combined exposure by convolving the Swift density map with the density map obtained from detection of 311 Swift GRBs with measured redshifts (their Fig.\,5). Since the 311-GRB density map already accounted for Swift's exposure (GRBs could only have been included in this distribution if Swift had detected them), the convolution improperly magnified the importance of Swift's exposure. 

Second, the extinction component of the exposure was not calculated independently of the data (as Swift's exposure had been). This means that any anisotropies inherent to the data were incorporated into the sky exposure calculation. \\
Third, in Sect.~\ref{sec:caveats} we briefly discuss the inadequacy of kernel based methods for the sky exposure function reconstruction in anisotropy studies.

\citet{Ukwatta:2016} also chose to compare the improperly-corrected density map to the data with only a 2D Kolmogorov--Smirnov test and the $n$th nearest neighbour test, excluding the more sensitive point-radius bootstrap method used by \citet{hhb14} and discussed in Sect.~\ref{sec:point}. One important advantage of the point-radius bootstrap method is that it randomly selects GRBs from sky regions according to the observed data, and thus follows the uncorrected density function rather than a uniform one.

As shown in Fig.~17 of \citet{Ukwatta:2016},  $\sim$60~GRBs per redshift bin are
required to exclude a chance alignment with a reasonably good
confidence. Although they estimated that it would take a decade to have a sufficiently large sample to make a definitive statement, the data we use in the present study already allows us to work with a sample size of~64. Nonetheless, we refrain from claiming here that the present sample size would be large enough already for drawing strong conclusions. We argue instead that -- based also on our Sect.~\ref{sec:Christiandata} showing the importance of a unified data set -- more data needs to be taken in a homogeneous way. We come back to the possibilities for increasing the sample size in Sect.~\ref{sec:observer}.

While we cannot completely exclude the possibility that sky exposure and galactic extinction might be responsible for the GRB anisotropy (we discuss this possibility further in Sect.~\ref{sec:skyexposure}), our results suggest that this is not the case, and the quantitative analysis presented by \citet{Ukwatta:2016} is not convincing in its current form. We therefore assume that the anisotropy is not caused by instrumental biases, and this leads us to propose several physical explanations as follows (Sect.~\ref{sec:screening}--\ref{sec:wave}). 

%--------------------------

\section{Discussion: the largest structure in the Universe?}\label{sec:discussion}

\subsection{The cosmological screening length}\label{sec:screening}

From certain perturbative theories of cosmology, a characteristic length is derived above which no cosmic structures can grow. It is called the screening length because, practically, the gravitational field gets damped above that distance due to the presence of gravitational objects. Below we elaborate on the prediction of such a characteristic screening scale from theory, including the summary of said theories and their consequences for our Universe's largest possible structures.

In a Universe that is \textit{perfectly} homogeneous and isotropic, the Einstein equations are typically solved by the means of a Friedmann–Lema{\^{i}}tre–Robertson–Walker metric \citep{1972bookWeinb,1984bookWald}. However, the real Universe is obviously not \textit{perfectly} homogeneous and isotropic -- there are small-scale variations in the distribution of matter for example, as the existence of stars, galaxies and other objects confirm. (For a review on embedding inhomogeneities in the Universe see e.g. \citealp{Faraoni:2018}.) Since this deviation from perfect homogeneity and isotropy is small, it is a reasonable approach to solve the Einstein equations up to a first order perturbation around the exact Friedman--Lema{\^{i}}tre--Robertson--Walker metric. 
This way we may get a fairly realistic model of cosmology in which the fact that objects exist is accounted for.

Such perturbative theories of General Relativity have been derived ({\it e.g.,} \citealp{Eingorn:2016} and \citealp{Hahn:2016}) by perturbing the metric (as well as the energy-momentum tensor) in the Einstein equations, while neglecting higher-order terms. The resulting equation takes the form of a well-known linear partial differential equation called the inhomogeneous Helmholtz equation \citep{1980PhRvD..22.1882B,1990CQGra...7.1169S,1999ASIC..541....1E}.

The solution of the inhomogeneous Helmholtz equation can be computed by a convolution with the fundamental solution (Green function) which, for this equation, has  the form of a Yukawa potential. What concerns us here is solving the equation of motion of an arbitrary particle of the system under the influence of all other particles, because this provides the form the gravitational force takes. By computing this \citep[cf. {\it e.g.,} Eq.~(11) of][]{Canay:2020} it turns out that the forces  caused by the inhomogeneities scale as e$^{-r/\lambda}$~$\times$~$1/r^2$. Notice that this form contains an exponentially decaying factor. Such a decay means that gravity gets damped (or in other words, screened) above the characteristic length~$\lambda$. 

This characteristic (screening) length depends on the terms applied in the initial perturbation of the metric and of the energy-momentum tensor in the Einstein equation. For example, \citet{Eingorn:2016} used a so-called discrete cosmology where the perturbation added to the energy-momentum tensor describes discrete sources, that is, point-like gravitational masses. On the other hand, \citet{Hahn:2016} applied continuous matter sources with small enough density fluctuations. Realising that both approaches are valid but at different scales, \citet{Canay:2020} took the next step and combined these two into a unified theory. The screening length these authors derived from their unified approach can be expressed as some function of the Hubble parameter~$H$ and the scaling factor~$a_0$ (see their Eq.~(43)). For today's values of the cosmological parameters, the screening length is obtained to be 2.57~Gpc.

\citet{Canay:2020} points out that this screening length of 2.57~Gpc exceeds the diameter of the Giant GRB Ring (1.7~Gpc) and matches the size of the Hercules--Corona Borealis Great Wall (2$-$3~Gpc). Again, this theoretically derived length is a consequence of solving the Einstein equation when matter is \textit{slightly} non-homogeneous and non-isotropic, and its quoted value has been obtained by unifying two perturbative approaches to the same problem. 

The theoretical calculations we summarised here are, of course, not free of assumptions -- starting with the fact that they all neglect higher order terms in the equations. Also, tensor perturbations of the metric (those associated with gravitational waves for example) are omitted due to their sources being way below accuracy limits of the present problem. Still, the formalism explained above is widely used in modern cosmological simulations. 

If the value that \citet{Canay:2020} derives for the cosmological screening length is valid, this would indeed be an interesting coincidence with the size of the Hercules--Corona Borealis Great Wall. To paraphrase these authors: provided that this colossal structure indeed exists, in light of their theoretical predictions for the screening length the Great Wall may be called not just ``the largest observed'' but simply ``the largest'' in the Universe.

\subsection{A direction-dependent Hubble constant?}\label{sec:Hubble}

While not reaching the scales investigated by perturbative cosmology, nor the scales traced out by GRBs, recent results from galaxy cluster analyses probing cosmic isotropy out to $z\lesssim 0.3$  are worth mentioning here \citep{Migkas:2020}. These authors find that, after excluding all possible alternative explanations for their measurements, an anisotropy of the Hubble constant, $H_0$, seems to persist. In other words, their measurements are consistent with the Hubble constant having different values at different celestial positions
(ranging between H$_0 =$~65 and 77~km/s/Mpc), as shown in their Fig.~23. 

We can only speculate if such an explanation could be behind the existence of the Hercules--Corona Borealis Great Wall too, albeit at much larger scales. For example, if the nature of dark-matter is such that it leads to different expansion rates for different directions in the Universe, that could possibly lead to measuring a non-constant H$_0$ for galaxy clusters, as well as observing large scale anisotropies in the GRB spatial distribution.

\subsection{A wave in star formation rate?}\label{sec:wave}

Astrophysical reasons for a clustering of GRBs are also worth to look at. 
Indeed, an excessive amount of gravitating matter would have an imprint on the cosmic microwave background due to the integrated Sachs--Wolfe effect \citep{Sachs:1967} but such an imprint is not seen.
Furthermore, analysing galaxies in the Millennium Simulation \citep{Angulo:2012}, \citet[][]{Balazs:2015} demonstrated that the spatial distribution of galaxies with normal vs. high star formation rate is probably different (cf. their Fig.~6). 
Therefore, as \citet[]{Balazs:2015} also noted, there may be astrophysical reasons other than cosmology and gravity for such anisotropies as the Great Wall. We present such speculative reasons in Sects.~\ref{sec:waves} and \ref{sec:metallicity}, but for this we need to discuss the nature of GRB progenitors and the host
galaxies first.

\subsubsection{GRB progenitors and star formation}

According to one of the most commonly accepted progenitor theories \citep[][]{Yoon:2005,Woosley:2006} long-duration GRBs (LGRBs, with prompt emission duration longer than~2\,s) originate from massive 20$-$30~M$_{\odot}$ stars  \citep{Yoon:2006,Cantiello:2007,Dessart:2008,vanMarle:2008} that rotate fast and undergo chemically homogeneous stellar evolution \citep[][]{Szecsi:2015} and explode in the collapsar scenario \citep{Woosley:1993,MacFadyen:1999}. For other progenitor theories, see a review in
\citep[][]{Szecsi:2017long}.
In the scenario mentioned above, a metallicity of $\lesssim$~0.5~Z$_{\odot}$ is required for such a massive star to maintain its sufficiently fast rotational rate throughout its life and to thus form a LGRB. The reason for this metallicity dependence is that massive stars lose angular momentum in their stellar winds, the strength of which is a function of the metallicity. For a more detailed discussion of such line-driven stellar winds and the spectral properties of LGRB progenitors, see \citet{Kubatova:2019}.
Formation of such a progenitor thus should take place in a host galaxy with low metallicity locally or globally.

Short-duration GRBs (SGRBs) are associated with the merger of two neutron stars i.e. the remnants of massive stars, which merge after a time interval in the order of billion years from birth \citep{Blinnikov:1984,Eichler:1989,Berger:2014,Ruiz:2016,Szecsi:2017short,VignaGomez:2018}.  
Intermediate-duration GRBs are not well understood from the point of view of a proposed progenitor. Since the majority of GRBs are long-duration anyway (only 28 out of the 487 GRBs studied here, or 6\%, have durations, T$_{\rm 90}< 2$~s), we focus the following discussion mainly on LGRBs.

From the initial mass function \citep{Salpeter:1955,Riaz:2020} and the fact that the lifetime of a star depends strongly on its mass \citep{Kippenhahn:1990}, it follows that massive stars are rare. As a consequence, wherever we see quite massive stars like an LGRB progenitor, we can suppose there is active star formation going on in that place.
On the other hand, the time the SGRB progenitor binary neutron stars need to merge can be quite long, so SGRBs are not expected to trace star-forming activity.

According to various studies, the cosmic star formation rate density
peaked 3~to 3.5~billion years after the Big Bang
\citep{Strolger:2004,Langer:2006,Rafelski:2012,Madau:2014,Madau:2017,Neijssel:2019}.
Based on that one may also expect the LGRB rate peaking at around $2<z<3$. 

\subsubsection{Waves of star formation}\label{sec:waves}

Whatever the cosmic star formation history was like, the \textit{spatial distribution} of star formation rate does not have to be uniform or isotropic at any given time.
Indeed, a clustering of LGRBs can be, in principle, the consequence of an enhanced star formation activity around the given spatial region without any significant increase of matter density. Just like star formation in galactic spiral arms.

In that picture the Great Wall is either a result of a significant single fluctuation or a (propagating?) wave of star formation rate. The latter would slowly undulate over cosmic time, enhancing star formation in one spacial direction at a given $z$ and in another direction at another $z$. The existence of such waves is not excluded by our data set. Although they would be present at all redshifts (i.e. not only at $z \sim 2$), the fact that the global star formation rate is observed to peak at around  $z\sim$~2$-$3  may mean that at other redshifts we simply do not have a large enough GRB sample to detect the ups and downs of these supposed waves.

In short, the Great Wall may be an upturn of a cosmic wave of star formation rate propagating over space-time. If we had more data at all redshifts, we might be able to follow these waves as they slowly heave and fade as the Universe evolves. 

If such waves exist and are the reason for the clustering of GRBs in the Great Wall, we can estimate their wavelengths. Following \citet[][who computed the lifetime of another large cosmic structure, cf. their Sect.~4.3.2 and their Eq.~(11)]{Balazs:2015}, and using the observed distance and size of 1~Gpc and 2~Gpc of the Great Wall, we obtain a wavelength of $\sim$10$^9$~yr. 

These speculations assumed that the rate of the GRB progenitor formation follows the general star formation rate in a given volume of space-time. However,
as a result of the Swift GRB Host Galaxy Legacy Survey spanning $0.03 < z < 6.3$, \citet{Palmerio:2019} claims that LGRBs are not good tracers of star formation. This implies that our speculations about a fluctuation or a wave in star formation rate is way too simplistic.
Still these authors find an enhanced fraction of starburst galaxies amongst the LGRB hosts. Thus, our simplistic picture can be improved by discussing the role the host's metallicity may play in this process.

\subsubsection{The role of metallicity}\label{sec:metallicity}

Finding some LGRBs 
in metal-poor host galaxies  \citep{Vergani:2015,Japelj:2016,Perley:2016,Palmerio:2019} is in accordance with the expected low metallicity of the progenitor. While the exact connection of the host's metallicity to the progenitor's metallicity is not necessarily trivial (about this see for example \citealp{Metha:2020} 
%\citep{Metha:2020} 
who investigated the scenario of high-metallicity galaxies hosting low-metallicity regions where LGRBs form), also from theory there seems to be a link between LGRB progenitors and metal-poor dwarf galaxies as pointed out by \citet{Szecsi:2015b}. 

In light of all this, the simple picture from Sect.~\ref{sec:waves} about GRBs tracing star formation activity and thus possibly evidencing some kind of wave in cosmic star formation rate, should be refined. LGRBs seems to trace \textit{metal-poor} star formation activity. So if the speculative interpretation of the Great Wall as some kind of wave propagating through cosmic space and time is valid, this could mean a wave in \textit{low-metallicity star formation rate} instead.

Further research is needed to figure out the origin of such a wave. 
For example, 
we refer to \citet{Mocz:2019} who studied a possible formation scenario of the first galaxies, suggesting that the first stars formed along dark matter filaments. Such a process may result in large scale metallicity anisotropies. And these may in turn result in an enhanced low-metallicity star formation activity and thus in anisotropies of the LGRBs distribution.

\subsection{A side-effect of observational biases?}\label{sec:caveats}

Cosmological calculations and astrophysical speculations aside, it would be premature to exclude observational bias of some sort as a feasible explanation. Indeed, \citet{Ukwatta:2016} also concluded using all-sky density maps that it is probably the sky exposure and galactic extinction that causes the finding of the Great Wall. While we have raised some concerns (cf. Sect.~\ref{sec:ukwatta}) about the details of their method, we do agree that there are complex observational biases arising when measuring GRB positions and redshifts, and that ideally these should be properly excluded before any definitive statement is made about the Great Wall's existence.

The detection probability of the redshift-determined GRBs in the sky is a combined probability. It depends not only on the various trigger conditions of the space instruments but the corresponding optical follow-ups, too. These include {\it e.g.,} timing, telescope/instrument availability, observer interest, observatory latitudes, seasonal weather, and Galactic extinction.

Albeit in principle it would be possible to integrate the satellite's known field-of-view timeline ({\it e.g.,} Swift pointings, see Sect.~\ref{sec:skyexposure}) and from this simulate the triggering, a synthetic reconstruction of the sky exposure for all satellites involved in our sample falls well outside the scope of the current work. And even if this were done, one would still need to combine it with a reliable model for selection effects related to the ground-based follow-up, which itself depends on many factors, including the impact of galactic extinction on detection by ground-based telescopes as well as purely human factors. 
Indeed, observers need to be interested enough in any given burst to dedicate their time and instrument to follow-up observations. This brings a possibly crucial yet rarely discussed component into the already complex problem of biases, which we investigate further in Sect.~\ref{sec:observer}.

\subsubsection{On the sky exposure}\label{sec:skyexposure}

Simple models for the sky exposure of Swift and, thus, at least for the biases in determining celestial position (i.e. pointing) of GRBs do exist. \citet{SWIFT10..060}, \citet{IAU15}, \citet{Li2015} as well as \citet{Ukwatta:2016} reconstructed the empirical sky exposure function of a GRB-like angular point process using kernel based methods.

Unfortunately however, these reconstructions require a kernel size to be set {\em a priori}: the typically observed number of GRBs puts the actual smoothing filter size in the $15-40{}^\circ$ range. Since the typical variability scale for the optical galactic extinction is smaller than that (the angular distribution of GRB's with redshift clearly shows a thin galactic disk), kernel based methods are expected to smooth out the fine details of the real exposure function. 
The kernel acts as a low-pass filter for the Fourier transform, and since the power spectrum is the square of the Fourier transform the filtering effect is stronger on it. As the two-point correlation function is the inverse Fourier transform of the power spectrum, it is smoothed too -- approximately applying the kernel filtering twice. Therefore all point statistics which are based on the estimated sky exposure function get distorted as finer details below the kernel size are not followed. This high frequency suppression effect is barely visible directly as the Poisson sampling usually generates higher noise level, but any analysis using large number of Monte Carlo simulations will contain (and sometimes clearly show) it (cf. \citealt{IAU15} and \citealt{Li2015}). The constraining effects of angular cell size was demonstrated in the measurement of galactic extinction by \citet{hak97}.

Alternatively, the sky exposure function should be more precisely determined with non-parametric density estimators, like in \citet{SWIFT10..060}, although the Poisson noise originating from the small numbers seems to dominate the sky exposure function estimation in this simple Voronoi-based methods. These techniques should be further refined and developed for a better analysis in the future.

\subsubsection{Observers' bias and future prospects with THESEUS}\label{sec:observer}

\begin{figure}
	\includegraphics[height=\columnwidth,angle=270]{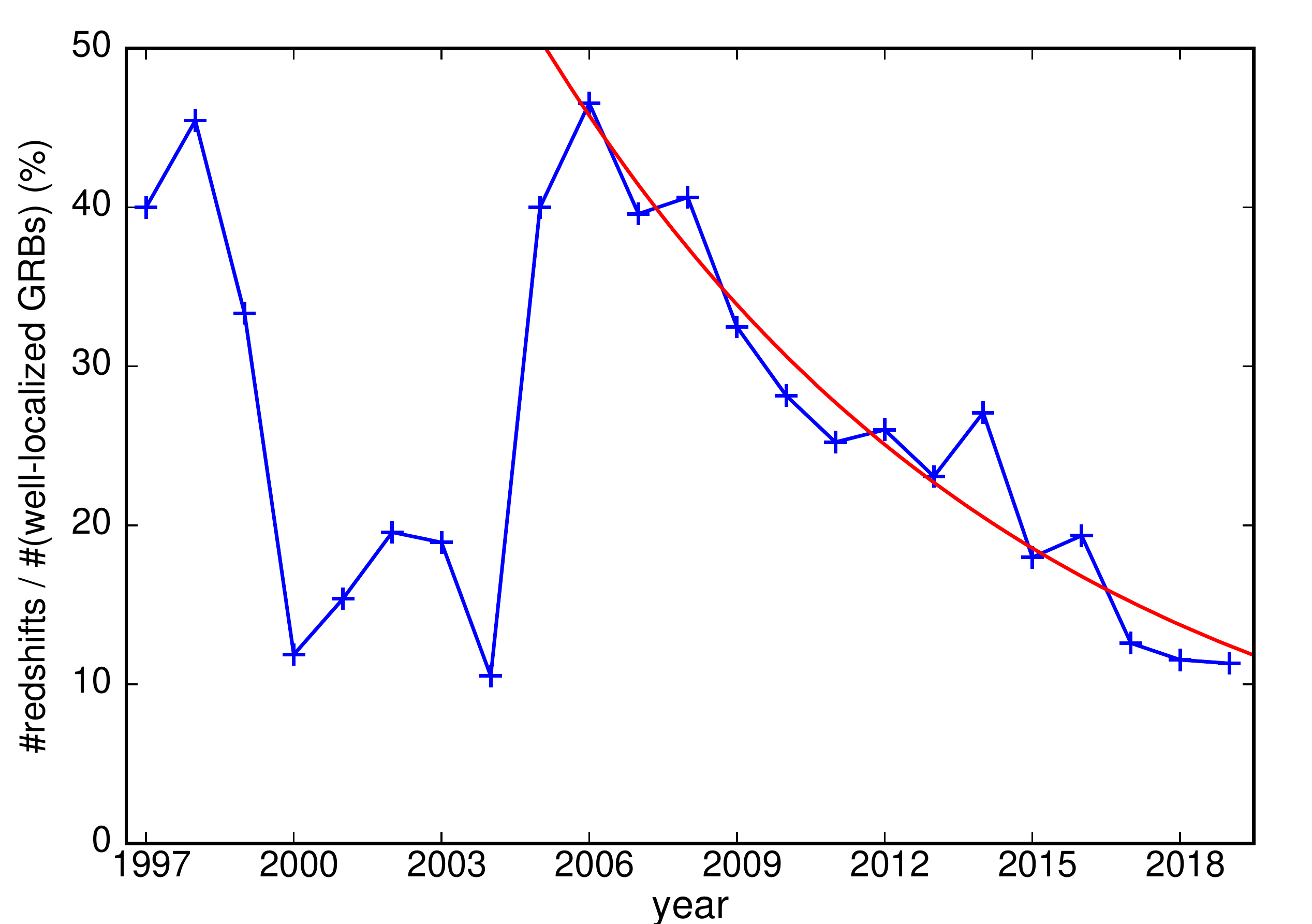}
    \caption{ Follow-up observations of the last two decades. The blue line shows the ratio of GRBs with redshift to all well-localised GRBs (i.e. localised within a few hours to days to less than $1^\circ$\,accuracy). The data are taken from Jochen Greiner's compilation (\url{http://www.mpe.mpg.de/~jcg/grbgen.html}). Since 2006 an approximately constant rate of well localised GRBs ($\approx 120$/year) are observed; however, the number of redshift measurements is clearly decreasing. This loss of interest in GRB redshift measurements amongst the observers can be well approximated with an exponential decay (red line). After 2006  only $\approx 90$\% of the previous year's redshift observations are obtained annually. If this trend continues, in 2026 we will observe less than 8 GRBs with redshift.}
    \label{fig:figjcg}
\end{figure}

Since Swift has been launched in 2004, the number of GRBs with well-determined sky position has been nearly constant ($\sim$120/year). However, the number of GRBs followed up by optical telescopes on the ground, has been continuously declining since: while it used to be $\sim$44/year in 2008, it was only $\sim$15/year in 2015 (see Fig.~\ref{fig:figjcg}). This decline has been consistent, showing a year-to-year decrease of $\sim$10\%\footnote{We used the publicly available data set of Jochen Greiner. \url{http://www.mpe.mpg.de/~jcg/grbgen.html}}.

This implies that, unless there is an extra feature raising interest -- such as an unexpectedly large redshift value indicated by the spacecraft's optical detector -- observers on the ground are less and less likely to dedicate their resources for an `average' (not so interesting) GRB. Indeed, it seems that if early optical afterglow detection (and sub-arc minute pointing) is done with Swift's UltraViolet and Optical Telescope, this vastly increases the chance that a ground-based follow-up and measuring of the spectrum of the afterglow would happen. For instance, for redshifts $z<1$ it enhances the chance by more than 60\%. 

From this we conclude two things. First, we believe this loss creates a very strong argument for building optical/UV/IR telescopes on board of upcoming gamma-satellites. The THESEUS mission \citep{Amati:2018,Stratta:2018}, for example, is currently being designed to host an Infrared Telescope \citep{Gotz:2018}. In light of the facts we report above, the importance of such a mission cannot be emphasised enough, both for motivating ground-based observers to follow-up interesting GRBs and for providing estimated redshifts for a large number of GRBs out to the epochs of the First Stars. Indeed, THESEUS will be essential for the future of studying cosmic isotropy with GRBs.
Second, the current sample of 487 GRBs with well-determined redshift may be biased by observer interest in a way that has previously not been accounted for.
If so, this may mean that despite enormous efforts of several communities to detect and localise GRBs with an ever increasing precision, the current data may only allow the study of cosmic isotropy in a limited and preliminary way. Again, future space missions such as THESEUS (which is currently undergoing assessment and, upon getting a green light, has an expected launch date of 2032) can change this by providing large, homogeneous samples of GRB redshift measurements. When this happens, our results of a statistically significant clustering of GRBs may need to be revised.

\section{Summary \& conclusions}\label{sec:summary}

If the Hercules--Corona Borealis Great Wall is real (and not, for example, an observational artefact), it is the largest  structure known in the Universe. Here we offered arguments both for and against it being real, and speculated about its potential origin. Our arguments and speculations are based on careful statistical tests we performed, as well as on the extensive study of the related literature. The main arguments we have thus derived are summarised as follows.

Using an up-to-date data set of all GRBs with reliably known redshift (Sect.~\ref{sec:data}), we performed two statistical tests. The first one ($k{\mbox{th}}$ nearest-neighbour analysis, Sect.~\ref{sec:nearest}) showed us that the group containing the redshift range $1.6 < z < 2.1$ is the most anisotropic. For this group (GR4) there were several cases in the $k=6-20$ range where the real data’s Kolmogorov--Smirnov like $D$ distance was above the 99.75\% of the random ones. However, the fact that no stronger peak was found implies that in future work more sensitive statistical methods should be applied to check the viability of this anomaly, as the used Kolmogorov--Smirnov like method is not sensitive enough.

The second test we performed (point radius bootstrap method, Sect.~\ref{sec:point}) is practically the same test that had been used to establish the possibility of the Great Wall's existence previously \citep{hhb14,hbht15}, however now with significantly more data points. Here, for the most complete data set currently available, the point radius bootstrap method showed us that the clustering of GRBs in this region of the Universe is indeed statistically significant. 

We dedicated Sect.~\ref{sec:comments} to addressing 
\citet{Ukwatta:2016} and
\citet{2020MNRASCS}, two papers that came to a different conclusion from ours.
We emphasize that both the nearest-neighbour analysis and the point radius bootstrap method, as used here, are independent of angular selection effects, provided these angular biases are not redshift-dependent. Thus we believe that the results obtained here are valid in the case of angular sky exposure by Swift, various ground observers' selection effects and for galactic extinction.

Investigating the literature for possible explanations of large structure formation, we found the recent study of \citet{Canay:2020} and discussed it in Sect.~\ref{sec:screening} -- this study in fact served as one of our main motivations for carrying out the present work. The authors concluded that, from perturbative cosmological theories, an effective length exists above which gravity gets screened. They derived this value to be, coincidentally, consistent with the size of the GRB Great Wall. Since such an interesting explanation for the origin of the Great Wall has been published in the literature, we thought it was in place to supplement it with further arguments.
Accordingly, alternative explanations were offered and discussed here, too. Motivated by recent results of galaxy cluster measurements probing cosmic isotropy, we speculated about the possibility that the Hubble constant may not, in fact, be constant in every direction of the sky (Sect.~\ref{sec:Hubble}).

Beside cosmological explanations, we also offered astrophysical ones (Sects.~\ref{sec:wave}). In particular, we speculated about systematic variations of the star-forming activity over cosmic time, including the role of metallicity. As our data cannot exclude a wave-like propagation of the star formation rate (note that the Great Wall coincides in redshift with the observed peak of cosmic star formation history), such waves have been suggested as another speculative source of a clustering of GRBs in the Great Wall. 

Although not scientifically conclusive at any rate,  we have created a video showing the orthographic 3D representation of the 4D GRB Universe to present the community with a means to visualise the Great Wall amongst all GRBs with known redshift (cf. Fig.~\ref{fig:3D} and Sect.~\ref{sec:GRBUniverse}). In the presently published video the position and distance of the Great Wall have been marked, but we are glad to provide similar videos marking any position and distance values upon request.

A 35\% increase in the GRB sample size has allowed us to re-explore our hypothesis regarding the existence of the Hercules--Corona Borealis Great Wall. However, as we have demonstrated (Sect.~\ref{sec:caveats}), the window of opportunity created for the GRB community by Swift may be closing. Observer fatigue appears to be reducing the rate at which GRBs with known redshift are measured, thus making it harder for large-scale GRB isotropy studies to continue into the future.

We are hopeful that this problem will be resolved because
the proposed gamma-satellite mission THESEUS has been designed to continue collecting a uniform and homogeneous GRB dataset. Having an infrared telescope on-board, THESEUS can provide us with just the data we need to study large-scale universal structures using GRBs and to continue testing whether or not the Hercules--Corona Borealis Great Wall is indeed real. If it is, it may well be the largest observable structure in the Universe. We need THESEUS to decide.

\section*{Acknowledgements}

The authors thank the Hungarian TIP and TKP program for their support.  
D.Sz. is supported by the Alexander von Humboldt Foundation. 
The authors thank the anonymous referee for their useful suggestions which have significantly increased the quality of the work.

\section*{Data availability}

The data underlying this article are available in GRBOX : Gamma-Ray Burst Online Index. 
The datasets were derived from sources in the public domain: https://www.astro.caltech.edu/grbox/grbox.php.

%%%%%%%%%%%%%%%%%%%%%%%%%%%%%%%%%%%%%%%%%%%%%%%%%%



%\bibliographystyle{mnras}

%\begin{thebibliography}{99}

%\bibliographystyle{mnras} 
%\bibliography{horv20,References}

\begin{thebibliography}{}
\makeatletter
\relax
\def\mn@urlcharsother{\let\do\@makeother \do\$\do\&\do\#\do\^\do\_\do\%\do\~}
\def\mn@doi{\begingroup\mn@urlcharsother \@ifnextchar [ {\mn@doi@}
  {\mn@doi@[]}}
\def\mn@doi@[#1]#2{\def\@tempa{#1}\ifx\@tempa\@empty \href
  {http://dx.doi.org/#2} {doi:#2}\else \href {http://dx.doi.org/#2} {#1}\fi
  \endgroup}
\def\mn@eprint#1#2{\mn@eprint@#1:#2::\@nil}
\def\mn@eprint@arXiv#1{\href {http://arxiv.org/abs/#1} {{\tt arXiv:#1}}}
\def\mn@eprint@dblp#1{\href {http://dblp.uni-trier.de/rec/bibtex/#1.xml}
  {dblp:#1}}
\def\mn@eprint@#1:#2:#3:#4\@nil{\def\@tempa {#1}\def\@tempb {#2}\def\@tempc
  {#3}\ifx \@tempc \@empty \let \@tempc \@tempb \let \@tempb \@tempa \fi \ifx
  \@tempb \@empty \def\@tempb {arXiv}\fi \@ifundefined
  {mn@eprint@\@tempb}{\@tempb:\@tempc}{\expandafter \expandafter \csname
  mn@eprint@\@tempb\endcsname \expandafter{\@tempc}}}

\bibitem[\protect\citeauthoryear{{Amati} et~al.,}{{Amati}
  et~al.}{2018}]{Amati:2018}
{Amati} L.,  et~al., 2018, \mn@doi [Advances in Space Research]
  {10.1016/j.asr.2018.03.010}, \href
  {http://adsabs.harvard.edu/abs/2018AdSpR..62..191A} {62, 191}

\bibitem[\protect\citeauthoryear{{Andrade}, {Bengaly}, {Alcaniz}  \&
  {Capozziello}}{{Andrade} et~al.}{2019}]{2019MNRAS.490.4481And}
{Andrade} U.,  {Bengaly} C. A.~P.,  {Alcaniz} J.~S.,   {Capozziello} S.,  2019,
  \mn@doi [\mnras] {10.1093/mnras/stz2754}, \href
  {https://ui.adsabs.harvard.edu/abs/2019MNRAS.490.4481A} {490, 4481}

\bibitem[\protect\citeauthoryear{{Angulo}, {Springel}, {White}, {Jenkins},
  {Baugh}  \& {Frenk}}{{Angulo} et~al.}{2012}]{Angulo:2012}
{Angulo} R.~E.,  {Springel} V.,  {White} S.~D.~M.,  {Jenkins} A.,  {Baugh}
  C.~M.,   {Frenk} C.~S.,  2012, \mn@doi [\mnras]
  {10.1111/j.1365-2966.2012.21830.x}, \href
  {https://ui.adsabs.harvard.edu/abs/2012MNRAS.426.2046A} {426, 2046}

\bibitem[\protect\citeauthoryear{{Bagoly}, Bal\'azs, {Horv{\'a}th}, {R{\'a}cz},
  {T{\'o}th}  \& {Hakkila}}{{Bagoly} et~al.}{2014}]{SWIFT10..060}
{Bagoly} Z.,  Bal\'azs L.~G.,  {Horv{\'a}th} I.,  {R{\'a}cz} I.,  {T{\'o}th}
  L.~V.,   {Hakkila} J.,  2014, in PoS(SWIFT 10)060, {\em Swift: 10 Years of
  Discovery, Proceedings of Science}, Rome, Italy,2-5 December 2014. ,
  \mn@doi{10.22323/1.233.0060}

\bibitem[\protect\citeauthoryear{Bagoly, Horv\'ath, Hakkila  \& T\'oth}{Bagoly
  et~al.}{2015}]{IAU15}
Bagoly Z.,  Horv\'ath I.,  Hakkila J.,   T\'oth L.~V.,  2015, \mn@doi
  [Proceedings of the International Astronomical Union]
  {10.1017/S1743921315010182}, 11, 2–2

\bibitem[\protect\citeauthoryear{{Bal{\'a}zs}, {M{\'e}sz{\'a}ros}  \&
  {Horv{\'a}th}}{{Bal{\'a}zs} et~al.}{1998}]{bal98}
{Bal{\'a}zs} L.~G.,  {M{\'e}sz{\'a}ros} A.,   {Horv{\'a}th} I.,  1998, \aap,
  \href {http://adsabs.harvard.edu/abs/1998A%26A...339....1B} {339, 1}

\bibitem[\protect\citeauthoryear{{Bal{\'a}zs}, {M{\'e}sz{\'a}ros},
  {Horv{\'a}th}  \& {Vavrek}}{{Bal{\'a}zs} et~al.}{1999}]{bal99}
{Bal{\'a}zs} L.~G.,  {M{\'e}sz{\'a}ros} A.,  {Horv{\'a}th} I.,   {Vavrek} R.,
  1999, \aaps, \href {http://adsabs.harvard.edu/abs/1999A%26AS..138..417B}
  {138, 417}

\bibitem[\protect\citeauthoryear{{Bal{\'a}zs}, {Bagoly}, {Hakkila},
  {Horv{\'a}th}, {K{\'o}bori}, {R{\'a}cz}  \& {T{\'o}th}}{{Bal{\'a}zs}
  et~al.}{2015}]{Balazs:2015}
{Bal{\'a}zs} L.~G.,  {Bagoly} Z.,  {Hakkila} J.~E.,  {Horv{\'a}th} I.,
  {K{\'o}bori} J.,  {R{\'a}cz} I.~I.,   {T{\'o}th} L.~V.,  2015, \mnras, 452,
  2236

\bibitem[\protect\citeauthoryear{{Bal{\'a}zs}, {Rejt{\H{o}}}  \&
  {Tusn{\'a}dy}}{{Bal{\'a}zs} et~al.}{2018}]{Balazs:2018}
{Bal{\'a}zs} L.~G.,  {Rejt{\H{o}}} L.,   {Tusn{\'a}dy} G.,  2018, \mn@doi
  [\mnras] {10.1093/mnras/stx2550}, \href
  {https://ui.adsabs.harvard.edu/abs/2018MNRAS.473.3169B} {473, 3169}

\bibitem[\protect\citeauthoryear{{Bardeen}}{{Bardeen}}{1980}]{1980PhRvD..22.1882B}
{Bardeen} J.~M.,  1980, \mn@doi [\prd] {10.1103/PhysRevD.22.1882}, \href
  {https://ui.adsabs.harvard.edu/abs/1980PhRvD..22.1882B} {22, 1882}

\bibitem[\protect\citeauthoryear{{Berger}}{{Berger}}{2014}]{Berger:2014}
{Berger} E.,  2014, \mn@doi [Annual Review of Astronomy and Astrophysics]
  {10.1146/annurev-astro-081913-035926}, \href
  {http://adsabs.harvard.edu/abs/2014ARA%26A..52...43B} {52, 43}

\bibitem[\protect\citeauthoryear{{Blinnikov}, {Novikov}, {Perevodchikova}  \&
  {Polnarev}}{{Blinnikov} et~al.}{1984}]{Blinnikov:1984}
{Blinnikov} S.~I.,  {Novikov} I.~D.,  {Perevodchikova} T.~V.,   {Polnarev}
  A.~G.,  1984, Soviet Astronomy Letters, \href
  {http://adsabs.harvard.edu/abs/1984SvAL...10..177B} {10, 177}

\bibitem[\protect\citeauthoryear{{Briggs} et~al.,}{{Briggs}
  et~al.}{1996}]{Briggs96}
{Briggs} M.~S.,  et~al., 1996, \mn@doi [\apj] {10.1086/176867}, \href
  {http://adsabs.harvard.edu/abs/1996ApJ...459...40B} {459, 40}

\bibitem[\protect\citeauthoryear{{Canay} \& {Eingorn}}{{Canay} \&
  {Eingorn}}{2020}]{Canay:2020}
{Canay} E.,  {Eingorn} M.,  2020, arXiv e-prints, \href
  {https://ui.adsabs.harvard.edu/abs/2020arXiv200200437C} {p. arXiv:2002.00437}

\bibitem[\protect\citeauthoryear{{Cantiello}, {Yoon}, {Langer}  \&
  {Livio}}{{Cantiello} et~al.}{2007}]{Cantiello:2007}
{Cantiello} M.,  {Yoon} S.-C.,  {Langer} N.,   {Livio} M.,  2007, \aap, 465,
  L29

\bibitem[\protect\citeauthoryear{{Christian}}{{Christian}}{2020}]{2020MNRASCS}
{Christian} S.,  2020, \mn@doi [\mnras] {10.1093/mnras/staa1448}, \href
  {https://ui.adsabs.harvard.edu/abs/2020MNRAS.495.4291C} {495, 4291}

\bibitem[\protect\citeauthoryear{{Cline}, {Matthey}  \& {Otwinowski}}{{Cline}
  et~al.}{1999}]{Cline99}
{Cline} D.~B.,  {Matthey} C.,   {Otwinowski} S.,  1999, \mn@doi [\apj]
  {10.1086/308094}, \href {http://adsabs.harvard.edu/abs/1999ApJ...527..827C}
  {527, 827}

\bibitem[\protect\citeauthoryear{{Clowes}, {Harris}, {Raghunathan},
  {Campusano}, {S{\"o}chting}  \& {Graham}}{{Clowes} et~al.}{2013}]{clo12}
{Clowes} R.~G.,  {Harris} K.~A.,  {Raghunathan} S.,  {Campusano} L.~E.,
  {S{\"o}chting} I.~K.,   {Graham} M.~J.,  2013, \mn@doi [\mnras]
  {10.1093/mnras/sts497}, \href
  {http://adsabs.harvard.edu/abs/2013MNRAS.429.2910C} {429, 2910}

\bibitem[\protect\citeauthoryear{{Dessart}, {Burrows}, {Livne}  \&
  {Ott}}{{Dessart} et~al.}{2008}]{Dessart:2008}
{Dessart} L.,  {Burrows} A.,  {Livne} E.,   {Ott} C.~D.,  2008, \mn@doi [\apjl]
  {10.1086/527519}, \href
  {https://ui.adsabs.harvard.edu/abs/2008ApJ...673L..43D} {673, L43}

\bibitem[\protect\citeauthoryear{{Eichler}, {Livio}, {Piran}  \&
  {Schramm}}{{Eichler} et~al.}{1989}]{Eichler:1989}
{Eichler} D.,  {Livio} M.,  {Piran} T.,   {Schramm} D.~N.,  1989, \mn@doi
  [\nat] {10.1038/340126a0}, \href
  {http://adsabs.harvard.edu/abs/1989Natur.340..126E} {340, 126}

\bibitem[\protect\citeauthoryear{{Eingorn}}{{Eingorn}}{2016}]{Eingorn:2016}
{Eingorn} M.,  2016, \mn@doi [\apj] {10.3847/0004-637X/825/2/84}, \href
  {https://ui.adsabs.harvard.edu/abs/2016ApJ...825...84E} {825, 84}

\bibitem[\protect\citeauthoryear{{Ellis} \& {van Elst}}{{Ellis} \& {van
  Elst}}{1999}]{1999ASIC..541....1E}
{Ellis} G. F.~R.,  {van Elst} H.,  1999, in {Lachi{\`e}ze-Rey} M.,  ed.,  NATO
  Advanced Science Institutes (ASI) Series C Vol. 541, NATO Advanced Science
  Institutes (ASI) Series C. pp 1--116 (\mn@eprint {arXiv} {gr-qc/9812046})

\bibitem[\protect\citeauthoryear{{Faraoni}}{{Faraoni}}{2018}]{Faraoni:2018}
{Faraoni} V.,  2018, \mn@doi [Universe] {10.3390/universe4100109}, \href
  {https://ui.adsabs.harvard.edu/abs/2018Univ....4..109F} {4, 109}

\bibitem[\protect\citeauthoryear{{Fasano} \& {Franceschini}}{{Fasano} \&
  {Franceschini}}{1987}]{1987MNRASFF}
{Fasano} G.,  {Franceschini} A.,  1987, \mn@doi [\mnras]
  {10.1093/mnras/225.1.155}, \href
  {https://ui.adsabs.harvard.edu/abs/1987MNRAS.225..155F} {225, 155}

\bibitem[\protect\citeauthoryear{{Gott}, {Juri{\'c}}, {Schlegel}, {Hoyle},
  {Vogeley}, {Tegmark}, {Bahcall}  \& {Brinkmann}}{{Gott}
  et~al.}{2005}]{Gott05}
{Gott} III J.~R.,  {Juri{\'c}} M.,  {Schlegel} D.,  {Hoyle} F.,  {Vogeley} M.,
  {Tegmark} M.,  {Bahcall} N.,   {Brinkmann} J.,  2005, \mn@doi [\apj]
  {10.1086/428890}, \href {http://adsabs.harvard.edu/abs/2005ApJ...624..463G}
  {624, 463}

\bibitem[\protect\citeauthoryear{{G{\"o}tz} et~al.,}{{G{\"o}tz}
  et~al.}{2018}]{Gotz:2018}
{G{\"o}tz} D.,  et~al., 2018, \memsai, \href
  {https://ui.adsabs.harvard.edu/abs/2018MmSAI..89..148G} {89, 148}

\bibitem[\protect\citeauthoryear{{Hahn} \& {Paranjape}}{{Hahn} \&
  {Paranjape}}{2016}]{Hahn:2016}
{Hahn} O.,  {Paranjape} A.,  2016, \mn@doi [\prd] {10.1103/PhysRevD.94.083511},
  \href {https://ui.adsabs.harvard.edu/abs/2016PhRvD..94h3511H} {94, 083511}

\bibitem[\protect\citeauthoryear{{Hakkila}, {Myers}, {Stidham}  \&
  {Hartmann}}{{Hakkila} et~al.}{1997}]{hak97}
{Hakkila} J.,  {Myers} J.~M.,  {Stidham} B.~J.,   {Hartmann} D.~H.,  1997,
  \mn@doi [\aj] {10.1086/118624}, \href
  {http://adsabs.harvard.edu/abs/1997AJ....114.2043H} {114, 2043}

\bibitem[\protect\citeauthoryear{{Horv{\'a}th}, {Hakkila}  \&
  {Bagoly}}{{Horv{\'a}th} et~al.}{2014}]{hhb14}
{Horv{\'a}th} I.,  {Hakkila} J.,   {Bagoly} Z.,  2014, \mn@doi [\aap]
  {10.1051/0004-6361/201323020}, \href
  {http://adsabs.harvard.edu/abs/2014A%26A...561L..12H} {561, L12}

\bibitem[\protect\citeauthoryear{{Horv{\'a}th}, {Bagoly}, {Hakkila}  \&
  {T{\'o}th}}{{Horv{\'a}th} et~al.}{2015}]{hbht15}
{Horv{\'a}th} I.,  {Bagoly} Z.,  {Hakkila} J.,   {T{\'o}th} L.~V.,  2015,
  \mn@doi [\aap] {10.1051/0004-6361/201424829}, \href
  {http://adsabs.harvard.edu/abs/2015A%26A...584A..48H} {584, A48}

\bibitem[\protect\citeauthoryear{{Japelj} et~al.,}{{Japelj}
  et~al.}{2016}]{Japelj:2016}
{Japelj} J.,  et~al., 2016, \mn@doi [\aap] {10.1051/0004-6361/201628314}, \href
  {http://adsabs.harvard.edu/abs/2016A%26A...590A.129J} {590, A129}

\bibitem[\protect\citeauthoryear{Kippenhahn \& Weigert}{Kippenhahn \&
  Weigert}{1990}]{Kippenhahn:1990}
Kippenhahn R.,  Weigert A.,  1990, Stellar Structure and Evolution.
Springer

\bibitem[\protect\citeauthoryear{{Kub{\'a}tov{\'a}} et~al.,}{{Kub{\'a}tov{\'a}}
  et~al.}{2019}]{Kubatova:2019}
{Kub{\'a}tov{\'a}} B.,  et~al., 2019, \mn@doi [\aap]
  {10.1051/0004-6361/201834360}, \href
  {http://adsabs.harvard.edu/abs/2019A%26A...623A...8K} {623, A8}

\bibitem[\protect\citeauthoryear{{Langer} \& {Norman}}{{Langer} \&
  {Norman}}{2006}]{Langer:2006}
{Langer} N.,  {Norman} C.~A.,  2006, \apjl, 638, L63

\bibitem[\protect\citeauthoryear{{Li} \& {Lin}}{{Li} \& {Lin}}{2015}]{Li2015}
{Li} M.-H.,  {Lin} H.-N.,  2015, \mn@doi [\aap] {10.1051/0004-6361/201525736},
  \href {https://ui.adsabs.harvard.edu/abs/2015A&A...582A.111L} {582, A111}

\bibitem[\protect\citeauthoryear{{Litvin}, {Matveev}, {Mamedov}  \&
  {Orlov}}{{Litvin} et~al.}{2001}]{li01}
{Litvin} V.~F.,  {Matveev} S.~A.,  {Mamedov} S.~V.,   {Orlov} V.~V.,  2001,
  \mn@doi [Astronomy Letters] {10.1134/1.1381609}, \href
  {http://adsabs.harvard.edu/abs/2001AstL...27..416L} {27, 416}

\bibitem[\protect\citeauthoryear{MacFadyen \& Woosley}{MacFadyen \&
  Woosley}{1999}]{MacFadyen:1999}
MacFadyen A.~I.,  Woosley S.~E.,  1999, ApJ, 524, 262

\bibitem[\protect\citeauthoryear{{Madau} \& {Dickinson}}{{Madau} \&
  {Dickinson}}{2014}]{Madau:2014}
{Madau} P.,  {Dickinson} M.,  2014, \mn@doi [\araa]
  {10.1146/annurev-astro-081811-125615}, \href
  {https://ui.adsabs.harvard.edu/abs/2014ARA&A..52..415M} {52, 415}

\bibitem[\protect\citeauthoryear{{Madau} \& {Fragos}}{{Madau} \&
  {Fragos}}{2017}]{Madau:2017}
{Madau} P.,  {Fragos} T.,  2017, \mn@doi [\apj] {10.3847/1538-4357/aa6af9},
  \href {https://ui.adsabs.harvard.edu/abs/2017ApJ...840...39M} {840, 39}

\bibitem[\protect\citeauthoryear{{Magliocchetti}, {Ghirlanda}  \&
  {Celotti}}{{Magliocchetti} et~al.}{2003}]{mgc03}
{Magliocchetti} M.,  {Ghirlanda} G.,   {Celotti} A.,  2003, \mn@doi [\mnras]
  {10.1046/j.1365-8711.2003.06657.x}, \href
  {http://adsabs.harvard.edu/abs/2003MNRAS.343..255M} {343, 255}

\bibitem[\protect\citeauthoryear{{M{\'e}sz{\'a}ros}, {Bagoly}, {Horv{\'a}th},
  {Bal{\'a}zs}  \& {Vavrek}}{{M{\'e}sz{\'a}ros} et~al.}{2000}]{mesz00}
{M{\'e}sz{\'a}ros} A.,  {Bagoly} Z.,  {Horv{\'a}th} I.,  {Bal{\'a}zs} L.~G.,
  {Vavrek} R.,  2000, \mn@doi [\apj] {10.1086/309193}, \href
  {http://adsabs.harvard.edu/abs/2000ApJ...539...98M} {539, 98}

\bibitem[\protect\citeauthoryear{{Metha} \& {Trenti}}{{Metha} \&
  {Trenti}}{2020}]{Metha:2020}
{Metha} B.,  {Trenti} M.,  2020, \mn@doi [\mnras] {10.1093/mnras/staa1114},
  \href {https://ui.adsabs.harvard.edu/abs/2020MNRAS.495..266M} {495, 266}

\bibitem[\protect\citeauthoryear{{Migkas}, {Schellenberger}, {Reiprich},
  {Pacaud}, {Ramos-Ceja}  \& {Lovisari}}{{Migkas} et~al.}{2020}]{Migkas:2020}
{Migkas} K.,  {Schellenberger} G.,  {Reiprich} T.~H.,  {Pacaud} F.,
  {Ramos-Ceja} M.~E.,   {Lovisari} L.,  2020, \mn@doi [\aap]
  {10.1051/0004-6361/201936602}, \href
  {https://ui.adsabs.harvard.edu/abs/2020A&A...636A..15M} {636, A15}

\bibitem[\protect\citeauthoryear{{Mocz} et~al.,}{{Mocz}
  et~al.}{2019}]{Mocz:2019}
{Mocz} P.,  et~al., 2019, \mn@doi [\prl] {10.1103/PhysRevLett.123.141301},
  \href {https://ui.adsabs.harvard.edu/abs/2019PhRvL.123n1301M} {123, 141301}

\bibitem[\protect\citeauthoryear{{Neijssel} et~al.,}{{Neijssel}
  et~al.}{2019}]{Neijssel:2019}
{Neijssel} C.~J.,  et~al., 2019, \mn@doi [\mnras] {10.1093/mnras/stz2840},
  \href {https://ui.adsabs.harvard.edu/abs/2019MNRAS.490.3740N} {490, 3740}

\bibitem[\protect\citeauthoryear{{Palmerio} et~al.,}{{Palmerio}
  et~al.}{2019}]{Palmerio:2019}
{Palmerio} J.~T.,  et~al., 2019, \mn@doi [\aap] {10.1051/0004-6361/201834179},
  \href {https://ui.adsabs.harvard.edu/abs/2019A&A...623A..26P} {623, A26}

\bibitem[\protect\citeauthoryear{{Peacock}}{{Peacock}}{1983}]{Peacock83}
{Peacock} J.~A.,  1983, \mnras, \href
  {http://adsabs.harvard.edu/abs/1983MNRAS.202..615P} {202, 615}

\bibitem[\protect\citeauthoryear{{Perley} et~al.,}{{Perley}
  et~al.}{2016}]{Perley:2016}
{Perley} D.~A.,  et~al., 2016, \mn@doi [\apj] {10.3847/0004-637X/817/1/8},
  \href {http://adsabs.harvard.edu/abs/2016ApJ...817....8P} {817, 8}

\bibitem[\protect\citeauthoryear{{Planck Collaboration} et~al.,}{{Planck
  Collaboration} et~al.}{2018}]{Planck:2018}
{Planck Collaboration} et~al., 2018, arXiv e-prints, \href
  {https://ui.adsabs.harvard.edu/abs/2018arXiv180706209P} {p. arXiv:1807.06209}

\bibitem[\protect\citeauthoryear{{R{\'a}cz}, {Bal{\'a}zs}, {Bagoly}, {T{\'o}th}
   \& {Horv{\'a}th}}{{R{\'a}cz} et~al.}{2017}]{racz2017AIP}
{R{\'a}cz} I.~I.,  {Bal{\'a}zs} L.~G.,  {Bagoly} Z.,  {T{\'o}th} L.~V.,
  {Horv{\'a}th} I.,  2017, \mn@doi [AIP Conference Proceedings]
  {10.1063/1.4968995}, \href
  {https://ui.adsabs.harvard.edu/abs/2017AIPC.1792f001R} {1792, 060012}

\bibitem[\protect\citeauthoryear{{Rafelski}, {Wolfe}, {Prochaska}, {Neeleman}
  \& {Mendez}}{{Rafelski} et~al.}{2012}]{Rafelski:2012}
{Rafelski} M.,  {Wolfe} A.~M.,  {Prochaska} J.~X.,  {Neeleman} M.,   {Mendez}
  A.~J.,  2012, \mn@doi [\apj] {10.1088/0004-637X/755/2/89}, \href
  {https://ui.adsabs.harvard.edu/abs/2012ApJ...755...89R} {755, 89}

\bibitem[\protect\citeauthoryear{{Riaz}, {Schleicher}, {Vanaverbeke}  \&
  {Klessen}}{{Riaz} et~al.}{2020}]{Riaz:2020}
{Riaz} R.,  {Schleicher} D.~R.~G.,  {Vanaverbeke} S.,   {Klessen} R.~S.,  2020,
  \mn@doi [\mnras] {10.1093/mnras/staa787}, \href
  {https://ui.adsabs.harvard.edu/abs/2020MNRAS.494.1647R} {494, 1647}

\bibitem[\protect\citeauthoryear{{\v R\'\i pa} \& {Shafieloo}}{{\v R\'\i pa} \&
  {Shafieloo}}{2019}]{2019MNRAS.486.3027Ripa}
{\v R\'\i pa} J.,  {Shafieloo} A.,  2019, \mn@doi [\mnras]
  {10.1093/mnras/stz921}, \href
  {https://ui.adsabs.harvard.edu/abs/2019MNRAS.486.3027R} {486, 3027}

\bibitem[\protect\citeauthoryear{{Ruiz}, {Lang}, {Paschalidis}  \&
  {Shapiro}}{{Ruiz} et~al.}{2016}]{Ruiz:2016}
{Ruiz} M.,  {Lang} R.~N.,  {Paschalidis} V.,   {Shapiro} S.~L.,  2016, \mn@doi
  [\apjl] {10.3847/2041-8205/824/1/L6}, \href
  {http://adsabs.harvard.edu/abs/2016ApJ...824L...6R} {824, L6}

\bibitem[\protect\citeauthoryear{{Sachs} \& {Wolfe}}{{Sachs} \&
  {Wolfe}}{1967}]{Sachs:1967}
{Sachs} R.~K.,  {Wolfe} A.~M.,  1967, \mn@doi [\apj] {10.1086/148982}, \href
  {https://ui.adsabs.harvard.edu/abs/1967ApJ...147...73S} {147, 73}

\bibitem[\protect\citeauthoryear{Salpeter}{Salpeter}{1955}]{Salpeter:1955}
Salpeter E.,  1955, ApJ, 121, 161

\bibitem[\protect\citeauthoryear{{Scott} \& {Tout}}{{Scott} \&
  {Tout}}{1989}]{1989MNRAS109S}
{Scott} D.,  {Tout} C.~A.,  1989, \mn@doi [\mnras] {10.1093/mnras/241.2.109},
  \href {https://ui.adsabs.harvard.edu/abs/1989MNRAS.241..109S} {241, 109}

\bibitem[\protect\citeauthoryear{{Slechta} \& {Meszaros}}{{Slechta} \&
  {Meszaros}}{1997}]{1997ApSSMS}
{Slechta} M.,  {Meszaros} A.,  1997, \mn@doi [\apss] {10.1023/A:1000181217004},
  \href {https://ui.adsabs.harvard.edu/abs/1997Ap&SS.249....1S} {249, 1}

\bibitem[\protect\citeauthoryear{{Stewart}}{{Stewart}}{1990}]{1990CQGra...7.1169S}
{Stewart} J.~M.,  1990, \mn@doi [Classical and Quantum Gravity]
  {10.1088/0264-9381/7/7/013}, \href
  {https://ui.adsabs.harvard.edu/abs/1990CQGra...7.1169S} {7, 1169}

\bibitem[\protect\citeauthoryear{{Stratta} et~al.,}{{Stratta}
  et~al.}{2018}]{Stratta:2018}
{Stratta} G.,  et~al., 2018, \mn@doi [Advances in Space Research]
  {10.1016/j.asr.2018.04.013}, \href
  {http://adsabs.harvard.edu/abs/2018AdSpR..62..662S} {62, 662}

\bibitem[\protect\citeauthoryear{{Strolger} et~al.,}{{Strolger}
  et~al.}{2004}]{Strolger:2004}
{Strolger} L.-G.,  et~al., 2004, \mn@doi [\apj] {10.1086/422901}, \href
  {https://ui.adsabs.harvard.edu/abs/2004ApJ...613..200S} {613, 200}

\bibitem[\protect\citeauthoryear{{Sz{\'e}csi}}{{Sz{\'e}csi}}{2017a}]{Szecsi:2017short}
{Sz{\'e}csi} D.,  2017a, Contributions of the Astronomical Observatory Skalnate
  Pleso, \href {http://adsabs.harvard.edu/abs/2017CoSka..47..108S} {47, 108}

\bibitem[\protect\citeauthoryear{{Sz{\'e}csi}}{{Sz{\'e}csi}}{2017b}]{Szecsi:2017long}
{Sz{\'e}csi} D.,  2017b, Proceedings of Science, 2017 [arXiv:1710.05655], \href
  {http://adsabs.harvard.edu/abs/2017arXiv171005655S} {PoS(MULTIF2017)065}

\bibitem[\protect\citeauthoryear{{Sz{\'e}csi}, {Langer}, {Sanyal}, {Evans},
  {Bestenlehner}  \& {Raucq}}{{Sz{\'e}csi} et~al.}{2015a}]{Szecsi:2015b}
{Sz{\'e}csi} D.,  {Langer} N.,  {Sanyal} D.,  {Evans} C.~J.,  {Bestenlehner}
  J.~M.,   {Raucq} F.,  2015a, in {Hamann} W.-R.,  {Sander} A.,   {Todt} H.,
  eds, Proceedings of Wolf-Rayet Stars Workshop, Potsdam, Germany, 2015.
  p.189-192. pp 189--192

\bibitem[\protect\citeauthoryear{{Sz{\'e}csi}, {Langer}, {Yoon}, {Sanyal}, {de
  Mink}, {Evans}  \& {Dermine}}{{Sz{\'e}csi} et~al.}{2015b}]{Szecsi:2015}
{Sz{\'e}csi} D.,  {Langer} N.,  {Yoon} S.-C.,  {Sanyal} D.,  {de Mink} S.,
  {Evans} C.~J.,   {Dermine} T.,  2015b, \aap, 581, A15

\bibitem[\protect\citeauthoryear{{Tarnopolski}}{{Tarnopolski}}{2017}]{2017MNRASTarnop}
{Tarnopolski} M.,  2017, \mn@doi [\mnras] {10.1093/mnras/stx2356}, \href
  {https://ui.adsabs.harvard.edu/abs/2017MNRAS.472.4819T} {472, 4819}

\bibitem[\protect\citeauthoryear{{Ukwatta} \& {Wo{\'z}niak}}{{Ukwatta} \&
  {Wo{\'z}niak}}{2016}]{Ukwatta:2016}
{Ukwatta} T.~N.,  {Wo{\'z}niak} P.~R.,  2016, \mn@doi [\mnras]
  {10.1093/mnras/stv2350}, \href
  {https://ui.adsabs.harvard.edu/abs/2016MNRAS.455..703U} {455, 703}

\bibitem[\protect\citeauthoryear{{\donothing{Van}van}~Marle, {Langer}, {Yoon}
  \& {Garc{\'{\i}}a-Segura}}{{\donothing{Van}van}~Marle
  et~al.}{2008}]{vanMarle:2008}
{\donothing{Van}van}~Marle A.~J.,  {Langer} N.,  {Yoon} S.-C.,
  {Garc{\'{\i}}a-Segura} G.,  2008, \mn@doi [\aap]
  {10.1051/0004-6361:20078802}, \href
  {https://ui.adsabs.harvard.edu/abs/2008A%26A...478..769V} {478, 769}

\bibitem[\protect\citeauthoryear{{Vavrek}, {Bal{\'a}zs}, {M{\'e}sz{\'a}ros},
  {Horv{\'a}th}  \& {Bagoly}}{{Vavrek} et~al.}{2008}]{vbh08}
{Vavrek} R.,  {Bal{\'a}zs} L.~G.,  {M{\'e}sz{\'a}ros} A.,  {Horv{\'a}th} I.,
  {Bagoly} Z.,  2008, \mn@doi [\mnras] {10.1111/j.1365-2966.2008.13635.x},
  \href {http://adsabs.harvard.edu/abs/2008MNRAS.391.1741V} {391, 1741}

\bibitem[\protect\citeauthoryear{{Vergani} et~al.,}{{Vergani}
  et~al.}{2015}]{Vergani:2015}
{Vergani} S.~D.,  et~al., 2015, \mn@doi [\aap] {10.1051/0004-6361/201425013},
  \href {http://adsabs.harvard.edu/abs/2015A%26A...581A.102V} {581, A102}

\bibitem[\protect\citeauthoryear{{Vigna-G{\'o}mez} et~al.,}{{Vigna-G{\'o}mez}
  et~al.}{2018}]{VignaGomez:2018}
{Vigna-G{\'o}mez} A.,  et~al., 2018, \mn@doi [\mnras] {10.1093/mnras/sty2463},
  \href {http://adsabs.harvard.edu/abs/2018MNRAS.481.4009V} {481, 4009}

\bibitem[\protect\citeauthoryear{{Wald}}{{Wald}}{1984}]{1984bookWald}
{Wald} R.~M.,  1984, {General Relativity}.
University of Chicago Press

\bibitem[\protect\citeauthoryear{{Weinberg}}{{Weinberg}}{1972}]{1972bookWeinb}
{Weinberg} S.,  1972, {Gravitation and Cosmology: Principles and Applications
  of the General Theory of Relativity}.
John Wiley and Sons

\bibitem[\protect\citeauthoryear{Woosley}{Woosley}{1993}]{Woosley:1993}
Woosley S.~E.,  1993, ApJ, 405, 273

\bibitem[\protect\citeauthoryear{{Woosley} \& {Heger}}{{Woosley} \&
  {Heger}}{2006}]{Woosley:2006}
{Woosley} S.~E.,  {Heger} A.,  2006, \apj, 637, 914

\bibitem[\protect\citeauthoryear{Yoon \& Langer}{Yoon \&
  Langer}{2005}]{Yoon:2005}
Yoon S.-C.,  Langer N.,  2005, A\&A, 443, 643

\bibitem[\protect\citeauthoryear{{Yoon}, {Langer}  \& {Norman}}{{Yoon}
  et~al.}{2006}]{Yoon:2006}
{Yoon} S.-C.,  {Langer} N.,   {Norman} C.,  2006, \mn@doi [\aap]
  {10.1051/0004-6361:20065912}, \href
  {http://adsabs.harvard.edu/abs/2006A%26A...460..199Y} {460, 199}

\makeatother
\end{thebibliography}

%\end{thebibliography}

%\hyphenation{Post-Script Sprin-ger}

%%%%%%%%%%%%%%%%%%%%%%%%%%%%%%%%%%%%%%%%%%%%%%%%%%

%%%%%%%%%%%%%%%%% APPENDICES %%%%%%%%%%%%%%%%%%%%%

%%%%%%%%%%%%%%%%%%%%%%%%%%%%%%%%%%%%%%%%%%%%%%%%%%

% Don't change these lines
\bsp	% typesetting comment
\label{lastpage}
\end{document}